\documentclass[useAMS, usenatbib, usegraphicx]{mn2e}

\usepackage{url}

\title[$\gamma$-ray anisotropy from galactic substructure]{Can galactic dark matter substructure contribute to the cosmic gamma-ray anisotropy?}
\author[J.U. Lange and M.-C. Chu]{J.U. Lange$^{1, 2}$\thanks{Present address: Astronomy Department, Yale University, P.O. Box 208101, New Haven, CT 06520-8101, USA.}\thanks{E-mail: johannesulf.lange@yale.edu} and M.-C. Chu$^{2}$\\
$^{1}$Institut f\"ur Theoretische Physik, Universit\"at Heidelberg, Philosophenweg 16, 69120 Heidelberg, Germany\\
$^{2}$Department of Physics, The Chinese University of Hong Kong, Shatin, N.T., Hong Kong}
\begin{document}

\date{Accepted xxx. Received xxx}

\pagerange{\pageref{firstpage}--\pageref{lastpage}} \pubyear{xxx}

\maketitle

\label{firstpage}

\begin{abstract}
The annihilation of dark matter (DM) particles in the Milky Way can contribute to the diffuse gamma-ray background (DGRB). Due to the presence of substructures, this emission will appear anisotropic in a predictable way. We generate full-sky maps of the gamma-ray emission in galactic substructures from results of the high-resolution Via Lactea II N-body simulation of the Milky Way DM halo. We calculate the anisotropy pattern, taking into account different radial profiles of the DM distribution in substructures, cosmic variance, and the detection threshold, and compare it to the anisotropy in the DGRB observed by the \textit{Fermi} Large Area Telescope (LAT). By comparing the upper limits on the DM self-annihilation cross-section, $\langle \sigma v \rangle$, implied by the anisotropy to the intensity of the DGRB and detected sources in the LAT 2-yr Point Source Catalog, we find that galactic substructure cannot contribute to the anisotropies in the DGRB without strongly violating these observations. Our results challenge the perception that small-scale anisotropies in the DGRB can be used as a probe of DM annihilation in galactic subhaloes.
\end{abstract}


\section{Introduction}

In the $\Lambda$ cold dark matter ($\Lambda$CDM) model, $27$ per cent of the energy content of the Universe consists of a mass component with no optical counterpart. Thus, this component is called dark matter (DM) and understanding its nature is among the prime goals of modern cosmology. One of the most promising candidates for this matter component is the weakly interacting massive particle (WIMP), produced as a thermal relic in the early Universe. In addition to its gravitational influence on astrophysical objects, the nature of DM can be explored through direct and indirect detection methods. Assuming that DM is made up of Majorana particles implies that it could self-annihilate, which would give rise to standard model particles that can be detected with ordinary observatories. Thus, this method is called the indirect detection method and is the target of this work.

Gamma-rays, in contrast to charged cosmic rays, are not deflected by magnetic fields, and thus trace back to their origin. Searching for DM with gamma-rays is therefore among the most promising methods to probe the nature of DM. Typical targets for observations are astrophysical objects with high DM densities, such as the galactic centre \citep{DMA:GalacticCenter1, DMA:GalacticCenter2}, dwarf galaxies \citep{DMA:Dwarfs} or galaxy clusters \citep{DMA:Cluster}, and the diffuse gamma-ray background \citep[DGRB;][]{DMA:DGRBIntensity}. DM annihilation may contribute to the total intensity of the DGRB, but could also, due to its presence in gravitationally collapsed structures, exhibit a detectable anisotropy pattern \citep{DMA:DGRB:Anisotropy1, DMA:DGRB:Anisotropy2}. In particular, recent high-resolution N-body simulations of a Milky Way like halo, such as Via Lactea II \citep{ViaLacteaII} and Aquarius \citep{Aquarius}, have been used to study the population of DM substructure, called DM subhaloes, of the Milky Way halo. In addition to subhaloes currently resolved by simulations, it is expected that subhaloes can form on much smaller scales. This minimum self-bound halo mass is often assumed to be $10^{-6} \ M_\odot$, but its precise value depends on the particular DM model \citep{DMA:MultiSource}. This population of subhaloes can dramatically increase the degree of anisotropy of the gamma-ray emission from galactic DM \citep{DMA:DGRB:Anisotropy1}.

The excellent performance of the Large Area Telescope \citep[LAT;][]{Fermi-LAT} on board the \textit{Fermi} Gamma-Ray Space Telescope, in particular its unprecedented angular resolution, began a new era in gamma-ray astronomy and indirect DM searches. Recently, the \textit{Fermi}-LAT Collaboration has measured an excess in the DGRB anisotropy in the multipole range $155 \leq \ell < 505$ and in the energy range from $1$ to $50 \ \mathrm{GeV}$ \citep{DGRB:Anisotropy}. With common astrophysical sources, such as active galactic nuclei \citep{AGNs:DGRB:Anisotropy} and, in particular, blazars \citep{Blazars:DGRB:Anisotropy1, Blazars:DGRB:Anisotropy2}, being expected to produce the bulk of this excess, these measurements can be used to constrain the DM self-annihilation cross-section, $\langle \sigma v \rangle$ \citep{DMA:DGRB:Anisotropy:Constraints1, DMA:DGRB:Anisotropy:Constraints2, DMA:DGRB:Anisotropy:Constraints3}.

In this work, we aim to compare the limits implied by the anisotropy to those from the intensity and detected sources in the LAT 2-yr Point Source Catalog \citep[2FGL;][]{2FGL}. We describe in detail the impact of different DM density profiles and the role of the detection threshold for subhaloes on the anisotropy power spectrum (APS) of the galactic DM emission. The former one has recently been discussed in \cite{DMA:DGRB:Anisotropy:RadialDistribution}, though with a greatly increased mass resolution we get qualitatively very different results. The latter effect, the role of the detection threshold of individual subhaloes, has, at least to our knowledge, not been discussed for the sensitivity of the study in \cite{DGRB:Anisotropy}. However, we find it to have a huge impact on both the shape and the normalization of the APS from DM. Taking this into account, we find significantly weaker limits on the self-annihilation cross-section, $\langle \sigma v \rangle$, than those previously reported in \cite{DMA:DGRB:Anisotropy:Constraints2} and \cite{DMA:DGRB:Anisotropy:Constraints3}. Ultimately, our results show that a detectable anisotropy signature from galactic DM would severely violate observations of the intensity of the DGRB and detected sources.

This paper is organized as follows. In Section \ref{sec:Galactic Dark Matter Annihilation}, we review the DM annihilation process in the Milky Way and describe our particular method for generating the all-sky gamma-ray template maps. Section \ref{sec:Dark Matter Power Spectra} presents the results of our simulation, focusing on the role of different DM density profiles and cosmic variance. In Section \ref{sec:Comparison with Observations}, we compare our theoretical model with observations of the anisotropy, the intensity, and detected sources in the DGRB. We present a simplified model for our conclusions in Section \ref{sec:Analytical Model}. Finally, we summarize the main points of our work in Section \ref{sec:Conclusion}.

\section{Galactic Dark Matter Annihilation}
\label{sec:Galactic Dark Matter Annihilation}

The gamma-ray flux, the number of detected gamma-rays per unit angle, unit time, unit area, and energy, from DM annihilation is given by
\begin{equation}
I(\theta, \phi, E_\gamma) = \frac{dN_\gamma}{d\Omega dt dA dE} = \frac{1}{2} \frac{\langle \sigma v \rangle}{m_\chi^2} \int\limits_{\mathrm{l.o.s.}} \frac{d N_\gamma}{d E} (E_\gamma) \frac{\rho^2}{4 \pi} d\chi,
\label{eq:DMA:Intensity}
\end{equation}
where $\langle \sigma v \rangle$ is the self-annihilation cross-section, $m_\chi$ the WIMP mass, $\rho$ the DM density, $d N_\gamma/d E$ is the annihilation gamma-ray spectrum of DM and $\chi$ the comoving distance. The integral is performed along the line of sight of the coordinates $\theta$ and $\phi$. We neglect any velocity dependence of the self-annihilation cross-section and assume it to be constant (s-wave annihilation). Furthermore, we assume annihilation into $b\bar{b}$ throughout this paper. We calculate the annihilation spectrum, $d N_\gamma/d E$ , for the WIMP with \texttt{DARKSUSY}\footnote{\url{http://www.darksusy.org/}} \citep{DarkSUSY} and model the galactic DM density from the results of the high-resolution Via Lactea II simulation \citep{ViaLacteaII}.

\subsection{Smooth galactic main halo}

Following \cite{ViaLacteaII}, we parametrize the density of the smooth galactic main halo with an Einasto profile \citep{Einasto},
\begin{equation}
\rho(r) = \rho_s \exp \left( -\frac{2}{\alpha} \left[ \left( \frac{r}{r_s} \right)^\alpha - 1 \right] \right),
\end{equation}
with $\alpha = 0.170$, $r_s = 21.5 \ \mathrm{kpc}$, and $\rho_s = 6.57 \times 10^{-2} \ \mathrm{GeV} \ \mathrm{cm}^{-3}$. To build the gamma-ray map for the smooth galactic halo, we place the virtual observer at $r = 8 \ \mathrm{kpc}$ and calculate the intensity from DM annihilation out to a radius of $500 \ \mathrm{kpc}$ around the observer. This model gives a local DM density of $0.4 \ \mathrm{GeV} \ \mathrm{cm}^{-3}$, matching the value reported in \cite{DMDensity}.

\subsection{Resolved galactic subhaloes}

The $\Lambda$CDM model predicts a hierarchical structure formation process in which low-mass haloes form first. These small haloes could merge and become subhaloes of the main halo. This additional substructure boosts the overall annihilation rate due to $\langle \rho^2 \rangle \geq \langle \rho \rangle^2$ and could induce an anisotropy pattern. In this work, we use the publicly available subhalo catalogue of the Via Lactea II simulation\footnote{\url{http://www.ucolick.org/~diemand/vl/}} \citep{ViaLacteaII}. We only consider haloes with tidal mass $M_{\mathrm{tid}}$ bigger than $2 \times 10^6 \ M_\odot$ and a maximum circular velocity $V_{\mathrm{max}}$ greater than $4 \ \mathrm{km} \ \mathrm{s}^{-1}$ at $z = 0$ to avoid possible numerical effects. Altogether, roughly $3600$ subhaloes with this criterion are identified within a radius of $500 \ \mathrm{kpc}$ around the galactic centre. For every subhalo, the maximum circular velocity $V_{\mathrm{max}}$, the corresponding radius $r_{\mathrm{max}}$, the tidal mass $M_{\mathrm{tid}}$, the tidal radius $r_{\mathrm{tid}}$, and the position are given. As discussed in \cite{DMA:DGRB:Anisotropy2}, the values of $V_{\mathrm{max}}$ and $r_{\mathrm{max}}$ can be affected by the gravitational softening length $\varepsilon = 40 \ \mathrm{pc}$. Thus, we apply a correction to those values to take this into account (see \cite{DMA:DGRB:Anisotropy2}, for details). Besides the Einasto profile, we also consider the Navarro, Frenk \& White (NFW) profile \citep{NFW}:
\begin{equation}
\rho(r) = \frac{4 \rho_s}{(r / r_s) \left(1 + r / r_s \right)^2},
\end{equation}
where $\rho_s$ and $r_s$ are free parameters for each subhalo. We use the $V_{\mathrm{max}}$ and $r_{\mathrm{max}}$ values of each halo to fit the DM distribution to different density profiles. It is straightforward to show that $r_{\mathrm{max}} = 2.163 r_s$ and
\begin{equation}
\int \rho^2 dV = 1.227 \frac{V_{\mathrm{max}}^4}{r_{\mathrm{max}} G^2} \left( 1 - \frac{1}{(1 + c_{\mathrm{tid}})^3} \right)
\end{equation}
for the NFW profile, where $c_{\mathrm{tid}} = r_{\mathrm{tid}} / r_s$. Similarly, one finds for the different Einasto profiles $r_{\mathrm{max}} = 2.189, 2.043, 1.892 r_s$ and
\begin{equation}
\int \rho^2 dV = 1.611, 0.951, 0.647 \frac{V_{\mathrm{max}}^4}{r_{\mathrm{max}} G^2} \frac{\Gamma \left( \frac{3}{\alpha}, \frac{4 c_{\mathrm{tid}}^\alpha}{\alpha} \right)}{\Gamma \left( \frac{3}{\alpha} \right)},
\end{equation}
for $\alpha = 0.18, 0.30$, and $0.50$, respectively. Here, $\Gamma(x)$ is the gamma function and $\Gamma(x_1, x_2)$ is the lower incomplete gamma function. All the profiles have the same value at $r_{\mathrm{max}}$, but make very different predictions for the behaviour near the centre of the subhalo. For the Einasto profile with $\alpha = 0.50$, the density reaches a constant value shortly after $r_{\mathrm{max}}$. On the other hand, the Einasto profile with $\alpha = 0.18$ and the NFW profile predict much higher central densities. While the density of the main halo can be resolved down to radii smaller than $0.01 r_{\mathrm{max}}$ \citep{Aquarius}, the behaviour near the centre is much more uncertain for the subhaloes. Thus, the different profiles described here are a good representation of possible behaviours of the density profile inside $r_{\mathrm{max}}$. Besides the density near the core, the different profiles also give different values for the density outside of $r_{\mathrm{max}}$. These densities could, in principle, be resolved and certain profiles might give a bad fit to N-body simulations. But the annihilation rate outside of $r_{\mathrm{max}}$ is negligible for every profile. Thus, it is justified to use these different profiles for all radii, even if they might not accurately predict the density outside of $r_{\mathrm{max}}$.

\subsection{Unresolved galactic subhaloes}

\begin{figure}
  \includegraphics[width=\columnwidth]{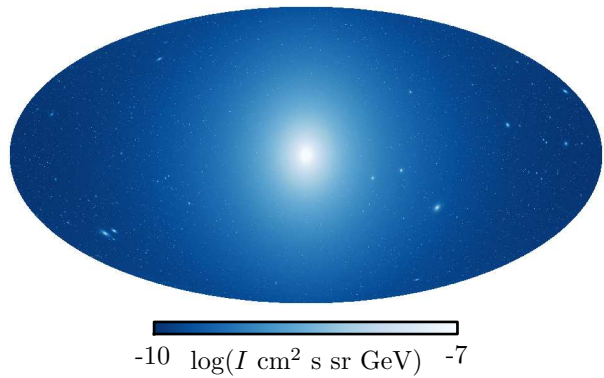}
  \caption{All-sky map of the gamma-ray emission at $E_\gamma = 4 \ \mathrm{GeV}$. We assume annihilation into $b \bar{b}$ and a WIMP mass, $m_\chi$, of $200 \ \mathrm{GeV}$.}
  \label{fig:GalacticFlux}
\end{figure}

Despite the excellent mass resolution in Via Lactea II, subhaloes can form at much smaller scales \citep{Microhalos}. It is thus important to quantify the possible effect of subhaloes below the resolution limit of the simulation. Here, a Monte Carlo approach is used to simulate a population of subhaloes down to a mass of $10^1 \ M_\odot$. Below this mass, the subhaloes are modelled with a substructure boost factor. We assume a power-law behaviour for the subhalo number density,
\begin{equation}
\frac{dN_{\mathrm{sub}}}{dM_{\mathrm{tid}}} = A \left( \frac{M_{\mathrm{tid}}}{M_\odot} \right)^{-\alpha}.
\label{eq:NumberDensity}
\end{equation}
With an index of $\alpha = 2$, this accurately describes the subhalo abundance of Via Lactea II \citep{ViaLacteaII}. The constant $A$ is fixed to $1.52 \times 10^{10} \ M_\odot^{-1}$ by the subhalo abundance in the range of $10^7 - 10^9 \ M_\odot$. An analytic fit for the concentration parameter, $c_{200}$, for the subhaloes in Via Lactea II is given in \cite{ImplicationsHigh-ResolutionSimulations}:
\begin{eqnarray}\nonumber
c_{200} (M_{\mathrm{tid}}, r) = & \left[ C_1 \left( \frac{M_{\mathrm{tid}}}{M_\odot} \right)^{-\alpha_1} + C_2 \left( \frac{M_{\mathrm{tid}}}{M_\odot} \right)^{-\alpha_2} \right] \\
& \times \left( \frac{r}{r_{\mathrm{vir}}} \right)^{-\alpha_r},
\label{eq:Concentration}
\end{eqnarray}
where $c_{200} = r_{200} / r_s$ and $r_{200}$ is the radius at which the average density inside the halo is $200$ times the critical density of the Universe. $r_{\mathrm{vir}} = 402 \ \mathrm{kpc}$ is the virial radius of the galactic halo in Via Lactea II, $r$ the distance to the galactic centre, $\alpha_r = 0.286$, $C_1 = 119.75$, $C_2 = -85.16$, $\alpha_1 = 0.012$, and $\alpha_2 = 0.0026$. Thus, this model predicts higher concentrations of subhaloes close to the galactic centre. Additionally, a Gaussian scatter of $\Delta \log_{10} c = 0.14$ is applied to the concentration parameter in the Monte Carlo simulation \citep{PredictionsGLAST}. Tidal forces near the galactic centre will strip the outer layers of a subhalo. The tidal radius $r_{\mathrm{tid}}$ of a given subhalo can be approximated by
\begin{equation}
r_{\mathrm{tid}} = \left[ \frac{M_{\mathrm{tid}}}{(2 - d \ln M_{\mathrm{gal}}(r) / d \ln r )M_{\mathrm{gal}}(r)} \right]^{1/3} r,
\end{equation}
with $M_{\mathrm{gal}}(r)$ being the integrated galactic mass within a sphere of radius $r$ \citep{Aquarius}. The distance $r$ for each subhalo is randomly drawn from the distribution of massive subhaloes,
\begin{equation}
N(< r) = N_{\mathrm{sub}} \frac{12 (r/500 \mathrm{kpc})^3}{1 + 11 (r/500 \mathrm{kpc})^2},
\end{equation}
where $N(<r)$ is the number of subhaloes inside a radius $r$ (compare \cite{ViaLacteaII:Subhalos}). Using this formula for all masses is motivated by the finding that the spatial number density is independent of the subhalo mass \citep{Aquarius}. Using the intrinsic relations for the NFW profile, $V_{\mathrm{max}}$ and $r_{\mathrm{max}}$ can be computed for each halo. For masses in the range of $10^1 - 2 \times 10^6 \ M_\odot$, all subhaloes generated are accepted. On the other hand, in the mass range from $2 \times 10^6$ to $10^7 \ M_\odot$, additional subhaloes are only accepted if they have a maximum circular velocity $V_{\mathrm{max}} < 4 \ \mathrm{km} \ \mathrm{s}^{-1}$ to prevent double counting of dense subhaloes. The resulting $r_{\mathrm{max}}$-$V_{\mathrm{max}}$ distribution smoothly extends the relation found in the simulation. Finally, the subhaloes below a mass of $10^1 \ M_\odot$ are taken into account using the substructure boost model by \cite{SubstructureBoost}. We consider a minimum self-bound halo mass of $10^{-6} \ M_\odot$. The luminosity of the unresolved substructure is assumed to follow an antibiased distribution,
\begin{equation}
\frac{dN_{\mathrm{sub}}}{dV} \propto r \rho_{\mathrm{gal}}(r),
\end{equation}
where $\rho_{\mathrm{gal}}(r)$ is the matter density of the smooth galactic halo. While these additional subhaloes can boost the total annihilation rate inside a halo significantly, we find that the intensity boost at the solar location is negligible, similar to the results in \cite{DMA:MultiSource}. The combined contribution of all galactic components is shown in Fig. \ref{fig:GalacticFlux}.

\section{Dark Matter Power Spectra}
\label{sec:Dark Matter Power Spectra}

Similar to \cite{DGRB:Anisotropy}, we define the APS, $C_\ell$, of an intensity map $\Psi(\theta, \phi)$ as
\begin{equation}
C_\ell = \frac{1}{f_{\mathrm{sky}} (2\ell + 1)} \sum\limits_{m=-\ell}^{\ell} |a_{\ell m}|^2,
\end{equation}
where $a_{\ell m}$ are the spherical harmonic coefficients and $f_{\mathrm{sky}}$ is the fraction of the sky used for the analysis. The coefficients are obtained by expanding the intensity map into spherical harmonics,
\begin{equation}
a_{\ell m} = \int \Psi(\theta, \phi) Y_{\ell m}^*(\theta, \phi) d\Omega,
\end{equation}
where the integral is performed over the unmasked part of the sky. Such a masking might be necessary because of strong galactic foreground emission. Additionally, we remove the monopole and dipole terms before the calculation. In this work, the power spectra are obtained with \texttt{HEALPIX}\footnote{\url{http://healpix.sourceforge.net}} \citep{HEALPix}, using a resolution of $N_{\mathrm{side}} = 512$.

\subsection{Role of the radial profile}

\begin{figure}
  \includegraphics[width=\columnwidth]{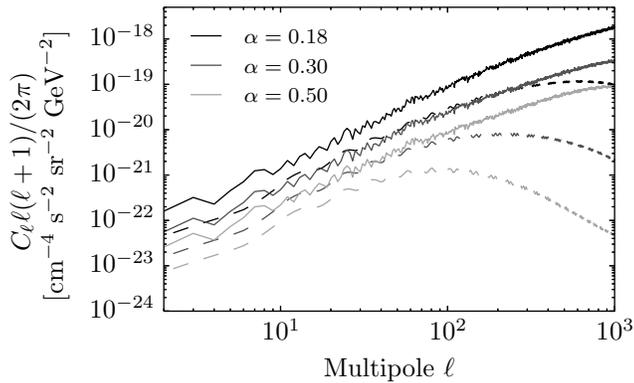}
  \caption{The anisotropy power spectra for galactic subhaloes with different Einasto profiles to describe the density distribution. The differential gamma-ray flux is measured at $E_\gamma = 4 \ \mathrm{GeV}$. We assume annihilation into $b \bar{b}$ and a WIMP mass, $m_\chi$, of $200 \ \mathrm{GeV}$. The gamma-ray power spectra are shown for subhaloes with a mass greater than $4 \times 10^8 \ M_\odot$ (dashed) and for subhaloes down to a mass of $10^1 \ M_\odot$ (solid).}
  \label{fig:ProfilesAnisotropy}
\end{figure}

The role of the density profile on the DM anisotropies has recently been discussed by \cite{DMA:DGRB:Anisotropy:RadialDistribution}. The authors used the results of the N-body counterparts of the MaGICC simulation suite to estimate the galactic DM signal. They used different parametrizations to fit the DM density profiles of the galactic smooth halo and its subhaloes. These were the NFW profile, the Einasto profile and the S\&M-profile \citep{SMProfile}. The latter one is given by
\begin{equation}
\rho(r) = \rho_0 \exp \left( - \lambda \left[ \ln \left( 1 + \frac{r}{R_\lambda} \right) \right]^2 \right),
\end{equation}
where $\rho_0$, $R_\lambda$, and $\lambda$ are free parameters. The analytical fits to these profiles were then used to construct all-sky gamma-ray maps. The authors show that the fluctuation angular power spectrum changes by `huge amounts' when using different density profiles to describe the simulation results. However, it is important to point out that the N-body simulation used by \cite{DMA:DGRB:Anisotropy:RadialDistribution} does only resolve subhaloes down to a minimum mass of $\sim 4 \times 10^8 \ M_\odot$. It is necessary to ask how this picture changes when analysing a broader mass range. We have simulated the galactic DM signal for the same subhalo population using the different Einasto profiles with $\alpha = 0.18$, $0.30$, and $0.50$ and without any mask. The anisotropy power spectra associated with these three different profiles are shown in Fig. \ref{fig:ProfilesAnisotropy}. When considering only subhaloes with a mass down to $4 \times 10^8 \ M_\odot$, we see huge differences between the density profiles. In particular, at a multipole of $\ell \sim 1000$, the difference is more than three orders of magnitude. The DM density profile affects both the shape and the normalization of the power spectrum, consistent with the results of \cite{DMA:DGRB:Anisotropy:RadialDistribution}. However, for the entire subhalo population down to a minimum mass of $10^1 \ M_\odot$ the differences decrease substantially. For cored profiles, such as Einasto with $\alpha = 0.50$, high-mass haloes do not inject strong anisotropies on small angular scales. In this case, the low-mass haloes provide the bulk of the anisotropy on small scales. In contrast, for cuspy profiles, like Einasto with $\alpha = 0.18$, most of the small-scale power already comes from high-mass haloes. Thus, taking only high-mass subhaloes into account, as done by \cite{DMA:DGRB:Anisotropy:RadialDistribution}, likely overestimates the effect of the radial profile. Especially the shape of the power spectrum is hardly affected by the choice of the radial profile.

\subsection{Cosmic variance}

\begin{figure}
  \includegraphics[width=\columnwidth]{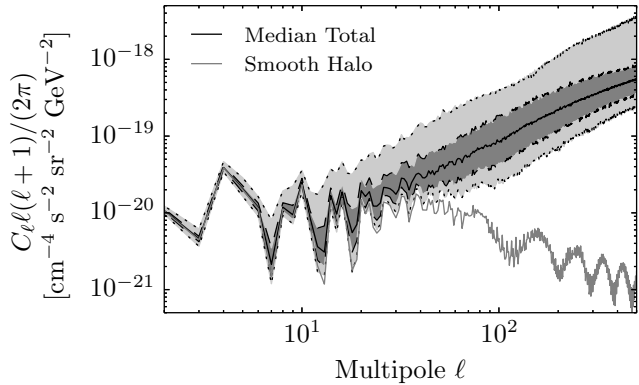}
  \caption{The power spectra of $100$ realizations of the galactic gamma-ray map. The particle physics parameters are the same as in Figs \ref{fig:GalacticFlux} and \ref{fig:ProfilesAnisotropy} and the NFW profile is used. The dark (light) grey areas show the $68$ per cent ($95$ per cent) containment ranges.}
  \label{fig:Variance}
\end{figure}

To test the variance of the DM anisotropy signal, we simulate $100$ realizations of the galactic gamma-ray map. For each realization of the map, we vary the position of the observer with respect to the galactic centre, while keeping a distance of $8 \ \mathrm{kpc}$ from the galactic centre. The DM density profile of the subhaloes is given by an NFW profile. Additionally, we mask low galactic latitudes ($|b| < 30^\circ$) and $2^\circ$ radius around each source in the 1FGL catalogue, as done in \cite{DGRB:Anisotropy}. This gives $f_{\mathrm{sky}} = 0.34$. We only consider subhaloes above a mass of $10^4 M_\odot$ because the contribution of lower mass subhaloes is negligible, as we will show later. The results of this analysis are shown in Fig. \ref{fig:Variance}. For the low multipoles ($\ell < 10$), the variance is negligible because the main contribution to the anisotropies comes from the smooth galactic halo. The anisotropy signal of the smooth halo, as shown by the grey line, does also have power on small angular scales. The features, however, are not intrinsic anisotropy features, but spurious remnants of the mask itself. For the multipole range of interest ($155 \leq \ell < 505$), the anisotropy signal is dominated by galactic substructure and, ultimately, affected by the low number of galactic high-mass subhaloes. The statistical uncertainty in terms of the $95$ per cent containment spans around one order of magnitude for $\ell = 100$. The variance decreases slightly for higher multipoles, which can be understood by the fact that rare, high-mass haloes have a smaller anisotropy contribution \citep{DMA:DGRB:Anisotropy1}. However, there is a high-anisotropy end for large multipoles that is due to very close encounters with dense nearby subhaloes. We also observe that different multipoles are highly correlated. Thus, differences in the realizations do not decrease significantly when averaging over a multipole range. However, for the remainder of this work, we will neglect this source of uncertainty. The reason is that the variance is caused by the most luminous subhaloes. As we will show in the next chapter, these subhaloes are too bright to cause a detectable anisotropy in the data of \cite{DGRB:Anisotropy} without being individually detected. We also note that these power spectra match the ones presented in \cite{DMA:DGRB:Anisotropy:Constraints1}, where Aquarius has been used to construct the all-sky maps, quite well. The anisotropies in our model are only slightly higher.

\section{Comparison with Observations}
\label{sec:Comparison with Observations}

\cite{DGRB:Anisotropy} have recently measured the anisotropy in the DGRB. They analysed the anisotropies in the energy ranges $1 - 2$, $2 - 5$, $5 - 10$, and $10 - 50 \ \mathrm{GeV}$ and in the multipole range $155 \leq \ell < 505$. An analysis of lower multipoles is hindered by contamination with galactic diffuse emission, while the upper limit is due to the angular reconstruction of the \textit{Fermi}-LAT. \cite{DGRB:Anisotropy} report an excess in angular power in all energy bins. Moreover, \cite{Blazars:DGRB:Anisotropy2} have estimated the contribution of unresolved blazars. Thus, one can use the residual anisotropy to place limits on the contribution of other unresolved source populations, e.g. DM annihilation. Recently, \cite{AGNs:DGRB:Anisotropy} have estimated the anisotropy originating from all unresolved AGNs combined. While the predicted anisotropies are slightly higher, they come with a higher uncertainty. Thus, they likely lead to a comparable residual anisotropy. In the following discussion, we will assume a WIMP mass of $m_\chi = 50 \ \mathrm{GeV}$, annihilation into $b \bar{b}$ and an NFW density profile for the galactic subhaloes. We will show how the masking of bright subhaloes affects the signature of DM annihilation as anisotropies in the DGRB. The results are qualitatively very similar for other particle physics parameters and DM density profiles.

\subsection{Detection threshold}

For the measurement of the gamma-ray anisotropy in \cite{DGRB:Anisotropy}, sources in the LAT 1-yr Point Source Catalog \citep[1FGL;][]{1FGL} have been masked. This catalogue contains several unassociated sources that could, in principle, be DM subhaloes. Thus, it is necessary to exclude DM subhaloes from the anisotropy calculation that exceed the threshold for individual detection. We follow \cite{Blazars:DGRB:Anisotropy2} and adopt a detection threshold $S_{\mathrm{th}}$ of
\begin{equation}
S_{\mathrm{th}} = 5 \times 10^{-10} \ \mathrm{cm}^{-2} \ \mathrm{s}^{-1}
\end{equation}
for $1 \ \mathrm{GeV} < E_\gamma < 100 \ \mathrm{GeV}$. \cite{Blazars:DGRB:Anisotropy2} showed that the detection threshold in this energy range is mostly independent of the underlying spectral shape in case of power-law emitters. Using the analytical estimate of \cite{1FGL}, we have verified that this is also the case for the gamma-ray spectrum of $\chi\chi \rightarrow b\bar{b}$. For this calculation, we used the estimate of the DGRB and the galactic diffuse emission presented in \cite{DGRB:Spectrum} and the P6 V3 Diffuse Instrument Response Functions of the \textit{Fermi}-LAT. For low DM masses, we find that the above value is a very good approximation, while the detection threshold for DM masses above $100 \ \mathrm{GeV}$ should be slightly lower. For simplicity and easy comparison, we will assume the above-mentioned detection threshold for all DM masses. We will later explore how our results are affected by different choices of the detection threshold.

There are, in principle, two important caveats to keep in mind when calculating the detection threshold according to \cite{1FGL}. First, this detection threshold is valid for point sources, whereas some subhaloes have a non-negligible angular extent. This could, in principle, increase the detection threshold. However, only the brightest subhaloes have a large angular size and the flux of these subhaloes is significantly above the threshold for a detectable anisotropy signal. All the remaining subhaloes have a negligible angular size, and thus can be treated as point sources. Secondly, the detection algorithm of the 1FGL assumes a power-law spectrum for the sources and is not optimized for DM sources. However, the 1FGL catalogue contains several pulsars that have a spectrum that is very similar to DM annihilation into $b\bar{b}$ \citep{DMA-PSR}. For galactic latitudes of $|b| >30^\circ$, the same region as for the anisotropy measurements, the catalogue contains six pulsars with $5 \times 10^{-10} \ \mathrm{cm}^{-2} \ \mathrm{s}^{-1} < S < 1 \times 10^{-9} \ \mathrm{cm}^{-2} \ \mathrm{s}^{-1}$ and nine with $1 \times 10^{-9} \ \mathrm{cm}^{-2} \ \mathrm{s}^{-1} < S < 1 \times 10^{-8} \ \mathrm{cm}^{-2} \ \mathrm{s}^{-1}$. If we assume that pulsars follow a power-law source count distribution with index $\gamma = 1.5$ (compare \cite{PSR:DGRB}), these numbers are perfectly compatible with a detection threshold of $S_{\mathrm{th}} = 5 \times 10^{-10} \ \mathrm{cm}^{-2} \ \mathrm{s}^{-1}$. Moreover, the more sensitive 2FGL catalogue reports no additional pulsars above $5 \times 10^{-10} \ \mathrm{cm}^{-2} \ \mathrm{s}^{-1}$. Finally, the spectra of most of the faint pulsars cannot be distinguished from power-law spectra. It is thus unlikely that the spectral shape degrades the detection efficiency of DM subhaloes dramatically.

Previously, \cite{DMA:DGRB:Anisotropy1} estimated the effect of the masking of bright subhaloes and found only a few detectable subhaloes for an observable anisotropy. However, the estimated detection threshold was much higher than the one of the 1FGL. \cite{DMA:DGRB:Anisotropy1} assumed a threshold of $2 \times 10^{-10} \ \mathrm{cm}^{-2} \ \mathrm{s}^{-1}$ for gamma-rays above $10 \ \mathrm{GeV}$ and 5 years of observation. However, the First \textit{Fermi}-LAT Catalog of Sources above $10 \ \mathrm{GeV}$ \citep[1FHL;][]{1FHL} reports a substantially lower detection threshold of $\sim 6 - 7 \times 10^{-11} \ \mathrm{cm}^{-2} \ \mathrm{s}^{-1}$ for only $3$ years of observation. Additionally, the 1FGL catalogue uses the entire energy range of the \textit{Fermi}-LAT, and thus has an even better detection efficiency. Moreover, \cite{DMA:DGRB:Anisotropy1} assumed that the anisotropy can be measured at $\ell \sim 10$, where we would have a much better signal-to-noise ratio than in the range $155 \leq \ell < 505$. Since this is not possible due to galactic foreground emission, a detectable anisotropy signal necessarily requires a higher self-annihilation cross-section, and thus more detectable subhaloes.

\subsection{Anisotropy}

\begin{figure}
  \includegraphics[width=\columnwidth]{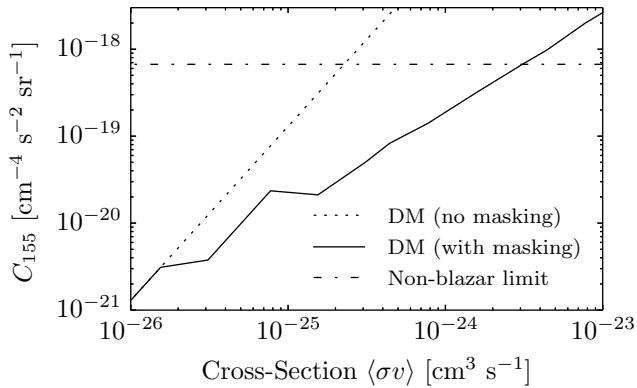}
  \caption{The APS from DM annihilation for $2 \ \mathrm{GeV} < E_\gamma < 5 \ \mathrm{GeV}$ and $\ell = 155$ as a function of the cross-section, $\langle \sigma v \rangle$. We assume annihilation into $b\bar{b}$ and $m_\chi = 50 \ \mathrm{GeV}$. The dot-dashed line denotes the $2\sigma$ upper limit on the non-blazar contribution \citep{Blazars:DGRB:Anisotropy2} in that energy bin.}
  \label{fig:Example_C155}
\end{figure}

\begin{figure}
  \includegraphics[width=\columnwidth]{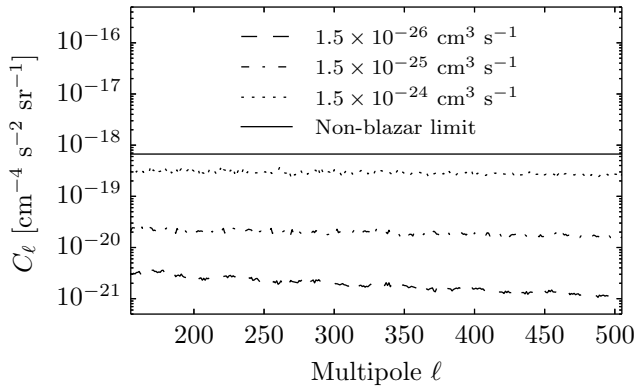}
  \caption{The APS from DM annihilation for $2 \ \mathrm{GeV} < E_\gamma < 5 \ \mathrm{GeV}$ for different values of the cross-section, $\langle \sigma v \rangle$. We exclude bright subhaloes that exceed the detection threshold, $S_{\mathrm{th}}$. A detectable APS from DM is completely Poisson-like, indicating that DM subhaloes in the DGRB are effectively point sources for the sensitivity of the Fermi-LAT.}
  \label{fig:Example_Cl}
\end{figure}

We exclude the contribution of the smooth galactic component and the subhaloes below $10^1 \ M_\odot$ for the anisotropy calculation because their contribution is negligible. Additionally, we choose a realization of the subhalo population where the anisotropy without masking bright subhaloes is near the median of the distribution shown in Fig. \ref{fig:Variance}. In Fig. \ref{fig:Example_C155}, we show how the power-spectrum coefficient at $\ell = 155$ in the energy range $2-5 \ \mathrm{GeV}$, which is the most sensitive energy range for this particle physics model, evolves as a function of the cross-section, $\langle \sigma v \rangle$. We also show the $2\sigma$ upper limit on the non-blazar contribution reported in \cite{Blazars:DGRB:Anisotropy2}. The dotted line, which scales as $C_{155} \propto \langle \sigma v \rangle^2$, shows how the anisotropy evolves if no bright subhaloes would be masked. Following the analysis of \cite{DMA:DGRB:Anisotropy:Constraints2}, this would yield a $95$ per cent confidence level (CL) upper limit of $\sim 2 \times 10^{-25}  \ \mathrm{cm}^3 \ \mathrm{s}^{-1}$ on the DM self-annihilation cross-section. However, masking bright subhaloes dramatically decreases the anisotropy from DM annihilation. Considering this effect, the power spectrum coefficient roughly scales linearly with the cross-section, $\langle \sigma v \rangle$, above the canonical cross-section value. Ultimately, the self-annihilation cross-section limit obtained is $\sim 3 \times 10^{-24}  \ \mathrm{cm}^3 \ \mathrm{s}^{-1}$, a factor of more than $10$ higher.

\begin{figure}
  \includegraphics[width=\columnwidth]{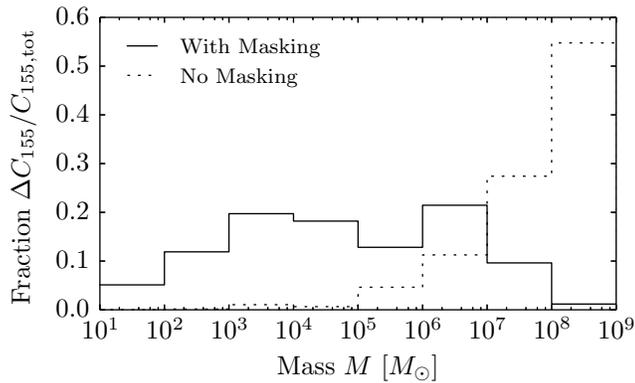}
  \caption{The contribution to the total anisotropy as a function of the subhalo mass. For this figure, we assumed a cross-section of $3 \times 10^{-24}  \ \mathrm{cm}^3 \ \mathrm{s}^{-1}$. While high-mass subhaloes dominate the anisotropy when masking is neglected, the contribution shifts considerably to lower mass subhaloes when the masking of bright subhaloes is applied.}
  \label{fig:Mass}
\end{figure}

Fig. \ref{fig:Example_Cl} shows the DM power spectra for different annihilation cross-sections in the multipole range $155 \leq \ell < 505$. For a small self-annihilation cross-section, with no subhaloes being masked, the angular power spectra are largely dominated by high-mass subhaloes with a considerable angular extent, as shown in Fig. \ref{fig:Mass}. This is demonstrated by an APS that decreases by a factor of $\sim 3$ towards higher multipoles. However, when increasing the annihilation cross-section, these subhaloes are masked and removed from the anisotropy calculation. The power spectrum is dominated by lower mass subhaloes and basically Poisson-like when the anisotropy from DM annihilation reaches the non-blazar anisotropy. This indicates that the anisotropy is created by unresolved sources with a negligible angular extent. Thus, a detectable angular signature from galactic DM annihilation is completely indistinguishable from astrophysical background sources, which are also expected to be dominated by a Poisson component \citep{AGNs:DGRB:Anisotropy}. This also implies that different DM density profiles, as discussed in Section \ref{sec:Dark Matter Power Spectra}, have basically no effect on the shape of the DM APS in the DGRB.

\subsection{Intensity}

\begin{figure}
  \includegraphics[width=\columnwidth]{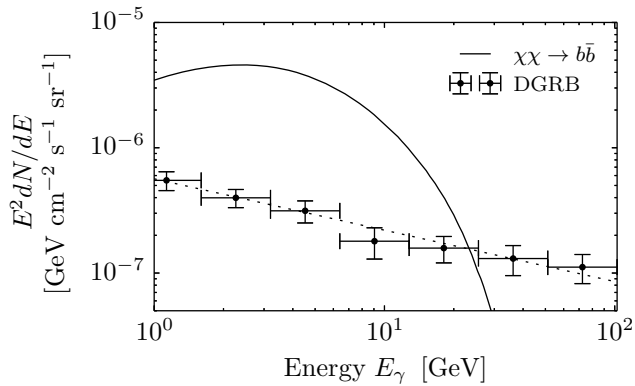}
  \caption{The intensity of the DGRB at high galactic latitudes ($|b| > 10^\circ$), as reported in \protect\cite{DGRB:Spectrum}, and the corresponding intensity from DM annihilation into $b \bar{b}$ implied by the upper limit from the anisotropy analysis.}
  \label{fig:Example_Spectrum}
\end{figure}

Let us compare the limits obtained from this anisotropy analysis to observations of individually detected sources and the intensity of the DGRB. Fig. \ref{fig:Example_Spectrum} shows the intensity from galactic DM annihilation implied by the upper limit of the anisotropy analysis and the spectrum of the DGRB \citep{DGRB:Spectrum}. The intensity for the galactic DM annihilation is averaged for high galactic latitudes, $|b| > 10^\circ$, as in the measurements of \cite{DGRB:Spectrum}. We see that a detectable angular signature from DM annihilation would imply that the intensity of DM overshoots the total intensity of the DGRB by roughly one order of magnitude. This is exacerbated by the fact the DGRB has multiple contributors and DM can at most contribute a small fraction to the total intensity of the DGRB \citep{DMA:DGRBIntensity}. Furthermore, most of the intensity from DM annihilation in our model comes from the smooth main halo. For a different substructure boost model with higher contributions from unresolved haloes, such as the one presented in \cite{HighSubstructureBoost}, this constraint would be even more severely violated.

\subsection{Detected Sources}

\begin{figure}
  \includegraphics[width=\columnwidth]{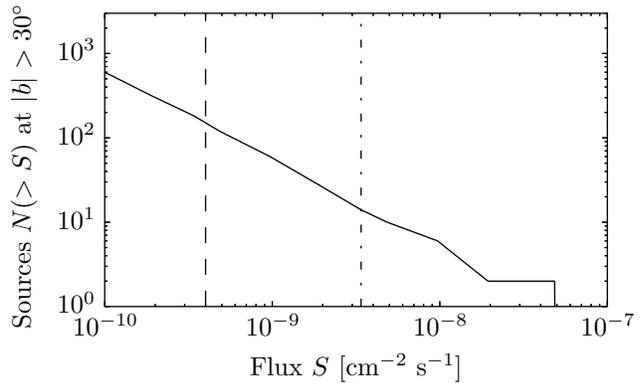}
  \caption{The subhalo luminosity function. The vertical dashed line shows the approximate detection threshold of the 2FGL catalogue and the vertical dot-dashed line the brightest flux of any observed DM subhalo candidate at high galactic latitudes ($|b| > 30^\circ$).}
  \label{fig:Example_Sources}
\end{figure}

Fig. \ref{fig:Example_Sources} shows the source count distribution of DM subhaloes with fluxes $S$ in the range of $1 - 100 \ \mathrm{GeV}$. We compare this to the detected sources in the 2FGL catalogue because, as the anisotropy measurement, it is based on 2 years of observation with the \textit{Fermi}-LAT. The vertical dashed line shows the detection threshold of the 2FGL of $F_{\mathrm{th}} \sim 4 \times 10^{-10} \ \mathrm{cm}^{-2} \ \mathrm{s}^{-1}$. For the upper limit implied by the anisotropy analysis, $\langle \sigma v \rangle = 3 \times 10^{-24}  \ \mathrm{cm}^3 \ \mathrm{s}^{-1}$, we find roughly $150$ detectable subhaloes at $|b| > 30^\circ$ in this catalogue. This violates observations of sources in three ways. First, this number strongly exceeds the $82$ unassociated sources reported in  \cite{DMA:SourceCount} that are located outside the galactic plane ($|b| > 30^\circ$), do not have a detectable time-variability and have no counterpart at other wavelengths. Additionally, only a small fraction of these DM subhalo candidates are actually compatible with the gamma-ray spectrum of $\chi \chi \rightarrow b \bar{b}$ with $m_\chi = 50 \ \mathrm{GeV}$. Secondly, several subhaloes would far exceed the detection threshold and reach luminosities of roughly $5 \times 10^{-8} \ \mathrm{cm}^{-2} \ \mathrm{s}^{-1}$. This exceeds the flux of the brightest subhalo candidates in the 2FGL that are compatible with a DM spectrum, as shown by the vertical dot-dashed line in the same figure, by more than one order of magnitude. Even more, \cite{DMA:SourceCount} only found four high-latitude subhalo candidates for that DM mass and annihilation channel above a flux of $10^{-9} \ \mathrm{cm}^{-2} \ \mathrm{s}^{-1}$, where the upper limit from the anisotropy analysis implies roughly $70$. Finally, these bright subhaloes would clearly be observed to have a non-negligible angular extent. However, no angular extension of any DM candidate source in the 2FGL has been detected so far \citep{DMA:SourceCount}. We conclude that a detectable anisotropy from DM annihilation would also severely violate observations of identified sources.

Let us finally remark that the theoretical model in \cite{DMA:SourceCount} produces many more bright subhaloes than our model. The upper limit of $\sim 2 \times 10^{-26}  \ \mathrm{cm}^3 \ \mathrm{s}^{-1}$ from subhalo searches reported in \cite{DMA:SourceCount} is not excluded for our theoretical predictions.

\subsection{Cross-section limits}

\begin{figure}
  \centering
  \includegraphics[width=\columnwidth]{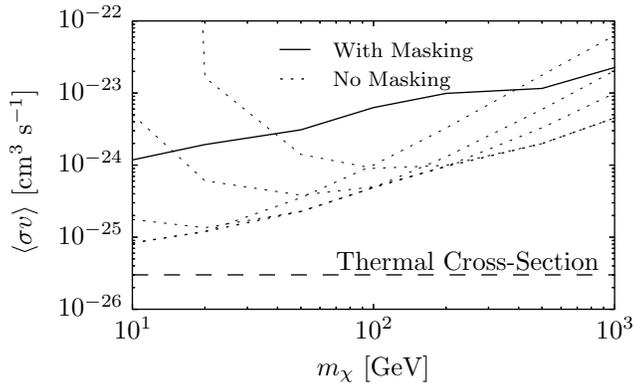}
  \caption{The $95$ per cent CL upper limits for the DM annihilation cross-section into $b\bar{b}$ implied by galactic substructure and the anisotropy of the DGRB.}
  \label{fig:Limits}
\end{figure}

Finally, the $95$ per cent CL upper limits on the annihilation cross-section into $b \bar{b}$ from the analysis of the anisotropies of galactic subhaloes are shown in Fig. \ref{fig:Limits}. For each mass, the upper limit is derived as the best upper limit in any of the four energy bins of the non-blazar anisotropy reported in \cite{Blazars:DGRB:Anisotropy2}. For reference, the upper limits derived when neglecting the masking of bright subhaloes is shown as a dotted line. We also show the individual limits from each energy bin as thin dashed lines. These constraints roughly match the limits presented in \cite{DMA:DGRB:Anisotropy:Constraints2}. However, when taking the masking of bright subhaloes into account, as shown by the solid line, the limits increase by a factor of $\sim 10$. The correction is slightly lower for very high WIMP masses. In this case, the annihilation of DM into $b\bar{b}$ produces very hard spectra. Thus, the detection threshold, $S_{\mathrm{th}}$, should be lower than for lighter DM particles and the effect of subhalo masking stronger than shown in Fig. \ref{fig:Limits}. The qualitative result is the same for all DM masses: A detectable anisotropy signal would severely violate observations of the intensity of the DGRB and individual sources.

\section{Analytical Model}
\label{sec:Analytical Model}

It is important to point out that the calculation of anisotropies, intensity, and detectable subhaloes from DM annihilation is subject to strong extrapolations beyond the range of N-body simulations, halo-to-halo variations of the substructure abundance or uncertainties in the mass of the Milky Way halo. Thus, one can ask whether the failure of the anisotropy analysis to provide competitive constraints on the annihilation cross-section is confined to Via Lactea II or our particular assumptions. We want to address this question with a simplified version of the analytical model presented in \cite{DMA:DGRB:Anisotropy1}. We choose this model for its simplicity and physical insight and refer the reader to \cite{DMA:DGRB:Anisotropy1} for a more sophisticated treatment.

The small-scale anisotropy from DM annihilation is dominated by the so-called one-subhalo term \citep{DMA:DGRB:Anisotropy1}. Generically, we can write its magnitude as
\begin{equation}
C_\ell = \int\limits_0^{S_{\mathrm{th}}} \frac{dN(>S)}{dS} \frac{K^2 S^2}{4 \pi} |\tilde{u}(\ell)|^2 dS,
\end{equation}
where $S$ is the flux of a source in a particular energy interval, $N(>S)$ the total number of sources above a certain flux $S$, $S_{\mathrm{th}}$ the detection threshold, $K$ a K-correction factor due to the fact that the luminosity and the anisotropy are not measured in the same energy range, and $|\tilde{u}(\ell)|$ a suppression factor due to angular extension. We will approximate this suppression factor with unity since we have already shown that sources below the detection threshold have a negligible angular extent. Assuming that the source count distribution $dN/dS$ follows a power-law relation, $dN/dS \propto S^{-\alpha}$, with index $\alpha$, the total small-scale anisotropy is
\begin{equation}
C_\ell = \frac{A(M_{\mathrm{MW}}, \langle \sigma v \rangle)}{4 \pi(3 - \alpha)} K^2  S_{\mathrm{th}}^{3 - \alpha},
\end{equation}
where $A(M_{\mathrm{MW}}, \langle \sigma v \rangle)$ is the normalization of the subhalo source count distribution that depends on the Milky Way mass and the self-annihilation cross-section. Conversely, the number of sources above a certain flux $S_*$ is
\begin{equation}
N(>S_*) = \int\limits_{S_*}^\infty \frac{dN(>S)}{dS} dS = \frac{A(M_{\mathrm{MW}}, \langle \sigma v \rangle)}{\alpha - 1} S_*^{1 - \alpha}.
\end{equation}
By changing the annihilation cross-section, we can fix the anisotropy level to the residual anisotropy $C_{\ell, 2\sigma}$. This then determines the total number of subhaloes above a certain flux $S_*$,
\begin{equation}
N(>S_*) = 4\pi C_{\ell, 2\sigma} \frac{3 - \alpha}{\alpha - 1} \frac{S_*^{1 - \alpha}}{K^2 S_{\mathrm{th}}^{3 - \alpha}}.
\end{equation}
Thus, the number of detected subhaloes for a detectable anisotropy is independent of the actual normalization of the substructure abundance. Thus, assuming self-similarity, it does also not depend on the Milky Way mass. Instead, it only depends on the index $\alpha$ of the source count distribution. For our model we get $\alpha \approx 2$. Using $C_{\ell, 2\sigma} = 6.7 \times 10^{-19} \ \mathrm{cm}^{-4} \ \mathrm{s}^{-2} \ \mathrm{sr}^{-1}$, $S_*$ = $10^{-9} \ \mathrm{cm}^{-2} \ \mathrm{s}^{-1}$, and $K = N_\gamma(2 - 5 \ \mathrm{GeV}) / N_\gamma(1 - 100 \ \mathrm{GeV}) = 0.36$, we get $N = 130 \times (1 - \sin 30^\circ) = 65$ at $|b| > 30^\circ$, in excellent agreement with the value shown in Fig. \ref{fig:Example_Sources}. We also see that the number of detected sources scales with $S_{\mathrm{th}}^{\alpha - 3} = S_{\mathrm{th}}^{-1}$. Thus, even for a very high detection threshold of $S_{\mathrm{th}} = 10^{-9} \ \mathrm{cm}^{-2} \ \mathrm{s}^{-1}$, the observations of individually detected sources in the 2FGL would still be violated.

\begin{figure}
  \includegraphics[width=\columnwidth]{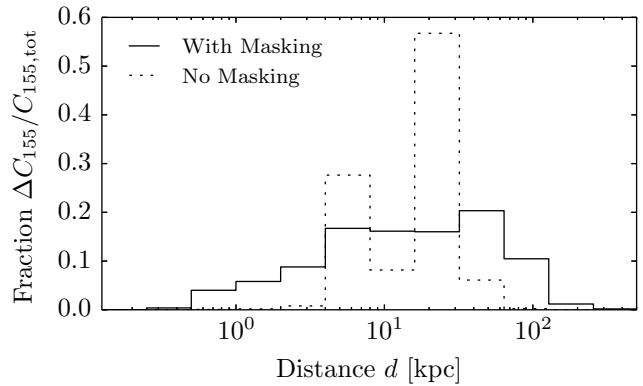}
  \caption{The contribution to the total anisotropy as a function of the distance $d$ of the subhalo to the observer. For this figure, we assumed a cross-section of $3 \times 10^{-24}  \ \mathrm{cm}^3 \ \mathrm{s}^{-1}$. The contribution of the unmasked subhalo population is dominated by outliers in luminosity. It is therefore not reliable.}
  \label{fig:Distance}
\end{figure}

In our simulation with Via Lactea II, the index $\alpha$ is approximately $2$. Interestingly, this index can be estimated through basic scaling relations. The absolute luminosity $L$ of a subhalo scales roughly as $c_{200}^3$. The concentration parameter $c_{200}$ is given by equation (\ref{eq:Concentration}) and scales as $c_{200} \propto R^{-\alpha_R}$, where $R$ is the distance from the galactic centre. Because most contributions to the anisotropy originate from subhaloes beyond a distance $d$ of $8 \ \mathrm{kpc}$ from the observer, as shown in Fig. \ref{fig:Distance}, we approximate the distance of the subhalo to the observer with the distance to the galactic centre, $d \sim R$. In this case, the apparent luminosity $S$ of a subhalo is proportional to $L R^{-2} \propto c_{200}^3 R^{-2} \propto R^{-(2 + 3 \alpha_R)}$. Finally, the number of subhaloes $N$ above some apparent luminosity $S$ scales as $N \propto R^{3} \propto S^{- 3 / (2 + 3 \alpha_R)}$. Here, we have neglected the decrease of subhalo abundance as a function of $R$. This would lower the index $\alpha$. Also note that in the above calculation, we have not assumed any mass-concentration relation or subhalo mass function. The index $\alpha$ does therefore not strongly depend on extrapolations to lower masses as long as $\alpha_R$ stays constant with mass. With $\alpha_R = 0.286$ for Via Lactea II this reproduces $\alpha \approx 3 / (2 + 3 \alpha_R) + 1 \approx 2.0$.

Would our results be qualitatively different for the Aquarius simulation? With $\alpha_R = 0.237$ \citep{ImplicationsHigh-ResolutionSimulations} this would yield a very similar result, $\alpha \approx 2.1$, and the number of subhaloes above $10^{-9} \ \mathrm{cm}^{-2} \ \mathrm{s}^{-1}$ would only decrease by $24$ per cent and still strongly violate observations of individually detected sources. Additionally, adopting a slope of $1.9$ for the subhalo number density in equation (\ref{eq:NumberDensity}), as done for Aquarius \citep{Aquarius}, would yield a higher contribution from distant, high-mass haloes and lower the index $\alpha$ again. To evade the constraints of individually detected sources, one would need a much higher value of $2.5$ or above. This could, in principle, be realized by an increasing concentration parameter or number density of subhaloes as a function of the distance from the galactic centre. But this strongly contradicts previous results from high-resolution simulations \citep{ViaLacteaII, Aquarius}.

\section{Conclusion}
\label{sec:Conclusion}

In this paper, we have predicted the DM annihilation APS from the high-resolution Via Lactea II N-body simulation and compared it to recent measurements of the gamma-ray anisotropy. We have first characterized the effect of different analytical density profiles and cosmic variance. In contrast to the results in \cite{DMA:DGRB:Anisotropy:RadialDistribution}, we found that different radial distributions only have a mild effect on the APS. We have then compared our predictions to the results in \cite{DGRB:Anisotropy} and showed that DM cannot contribute to the anisotropies of the DGRB without strongly violating observations from identified sources and the spectrum of the DGRB. For completeness, we show our limits on the DM cross-section, $\langle \sigma v \rangle$, into $b \bar{b}$ in Fig. \ref{fig:Limits}.

The apparent failure of anisotropies in the DGRB to place competitive constraints on DM annihilation in the Milky Way is based on a very general argument. The anisotropy in the DGRB is caused by a large number of astrophysical sources just below the detection threshold, $S_{\mathrm{th}}$. Since the signal-to-noise ratio for these measurements is not very high, DM would need to have a comparable number of sources, e.g. subhaloes, below the detection threshold to give a detectable anisotropy contribution. But, unless a very unusual behaviour of the source count distribution $dN/dS$ for DM subhaloes is invoked, this necessarily implies a large number of sources above the detection threshold.

Based on this argument, it seems unlikely that anisotropies can place any competitive constraints in the near future. While the precision of the anisotropy measurements will increase as \textit{Fermi} collects more data, so will the detection efficiency. Additionally, the blazar model in \cite{Blazars:DGRB:Anisotropy2} already explains almost all the anisotropies in the DGRB. Thus, it is unlikely that an advancement in the knowledge of the background sources will improve the anisotropy constraints substantially. However, we want to stress that cross-correlating the anisotropies with gravitational tracers \citep{DMA:Cross-Correlation} is a very promising way to probe extragalactic DM annihilation. We also do not intend to make estimates about this method at higher energy ranges \citep{DMA:Anisotropies:Cherenkov}. For the energy range of the \textit{Fermi}-LAT, however, galactic DM is likely to first be seen as identified sources or in the isotropic spectrum of the DGRB and not as small-scale anisotropies.

\section*{Acknowledgements}

We thank Kenny Ng and Shek Yeung for helpful discussions and comments on the manuscript. The work of J.U.L. was supported by a scholarship of the German Academic Exchange Service (DAAD). Some of the results in this paper have been derived using the \texttt{HEALPIX} \citep{HEALPix} package.

\label{lastpage}

\footnotesize{
\bibliographystyle{mn2e}
\bibliography{Bibliography}
}

\end{document}